\newif\ifarxiv 
\arxivtrue

\ifarxiv\documentclass[%
reprint,
superscriptaddress,
nofootinbib,
amsmath,amssymb,
aps,longbibliography
]{revtex4-1}
\else
\documentclass[reprint,
superscriptaddress,
amsmath,amssymb,
aps,longbibliography]{revtex4-1}
\fi
\usepackage{graphicx}
\usepackage[shortlabels]{enumitem}
\usepackage{mathtools}
\usepackage{fancyhdr}
\usepackage{nicefrac}
\usepackage{bbm}
\makeatletter

\usepackage{upgreek}
\newcommand{\changefont}{%
    \fontsize{9}{11}\selectfont
}
 \usepackage{makecell}
 \usepackage{placeins}
  \usepackage{afterpage}
\usepackage{mathrsfs}

\usepackage{amsfonts,amssymb,amsmath}            
\usepackage{ifthen}
\usepackage{amsthm, thm-restate}
\usepackage{dsfont}
\usepackage{float}

\usepackage[caption=false]{subfig}
\PassOptionsToPackage{hyphens}{url}
\usepackage[hyperfootnotes=false]{hyperref}
\usepackage{xcolor}
\definecolor{mylinkcolor}{rgb}{0,0,0.7} 
\hypersetup{unicode=true,%
  bookmarksnumbered=false,bookmarksopen=false,bookmarksopenlevel=1, %
  breaklinks=true,pdfborder={0 0 0},colorlinks=true}%
\hypersetup{%
  anchorcolor=mylinkcolor,citecolor=mylinkcolor, %
  filecolor=mylinkcolor,linkcolor=mylinkcolor, %
  menucolor=mylinkcolor,runcolor=mylinkcolor, %
  urlcolor=mylinkcolor}

\usepackage{tikz}
\usetikzlibrary{arrows}
\usetikzlibrary{arrows.meta}
\usepackage{rotating, xcolor}
\usetikzlibrary{shapes}
\usetikzlibrary{hobby}
\usetikzlibrary{decorations.markings}
\usetikzlibrary{arrows.meta}
\usetikzlibrary{snakes} 
\usepackage{tikz-cd}
\tikzset{degil/.style={
            decoration={markings,
            mark= at position 0.5 with {
                  \node[transform shape] (tempnode) {$\setminus$};
                  }
              },
              postaction={decorate}
}
}

\newcommand\longrsquigarrow{
\begin{tikzpicture}
\draw [decorate, decoration={zigzag, segment length=+6pt, amplitude=+.95pt,post length=+2pt}, arrows={-Classical TikZ Rightarrow}]  (0,0.1) -- (0.6,0.1); \draw[draw=none] (0,0)--(0.6,0);
\end{tikzpicture}
}

\newcommand\longlsquigarrow{
\begin{tikzpicture}
\draw [decorate, decoration={zigzag, segment length=+6pt, amplitude=+.95pt,post length=+2pt}, arrows={-Classical TikZ Rightarrow},rotate around={180:(0.3,0.1)}]  (0,0.1) -- (0.6,0.1); \draw[draw=none] (0,0)--(0.6,0);
\end{tikzpicture}
}

\makeatletter
\newcommand*{\da@rightarrow}{\mathchar"0\hexnumber@\symAMSa 4B }
\newcommand*{\da@leftarrow}{\mathchar"0\hexnumber@\symAMSa 4C }
\newcommand*{\xdashrightarrow}[2][]{%
  \mathrel{%
    \mathpalette{\da@xarrow{#1}{#2}{}\da@rightarrow{\,}{}}{}%
  }%
}
\newcommand{\xdashleftarrow}[2][]{%
  \mathrel{%
    \mathpalette{\da@xarrow{#1}{#2}\da@leftarrow{}{}{\,}}{}%
  }%
}
\newcommand*{\da@xarrow}[7]{%
  \sbox0{$\ifx#7\scriptstyle\scriptscriptstyle\else\scriptstyle\fi#5#1#6\m@th$}%
  \sbox2{$\ifx#7\scriptstyle\scriptscriptstyle\else\scriptstyle\fi#5#2#6\m@th$}%
  \sbox4{$#7\dabar@\m@th$}%
  \dimen@=\wd0 %
  \ifdim\wd2 >\dimen@
    \dimen@=\wd2 %
  \fi
  \count@=2 %
  \def\da@bars{\dabar@\dabar@}%
  \@whiledim\count@\wd4<\dimen@\do{%
    \advance\count@\@ne
    \expandafter\def\expandafter\da@bars\expandafter{%
      \da@bars
      \dabar@ 
    }%
  }%
  \mathrel{#3}%
  \mathrel{%
    \mathop{\da@bars}\limits
    \ifx\\#1\\%
    \else
      _{\copy0}%
    \fi
    \ifx\\#2\\%
    \else
      ^{\copy2}%
    \fi
  }%
  \mathrel{#4}%
}
\makeatother

\usepackage[nodayofweek]{datetime}
\newcommand{\comment}[1]{}

\newcommand{\cA}{\mathcal{A}}
\newcommand{\cB}{\mathcal{B}}
\newcommand{\cC}{\mathcal{C}}

\newcommand{\cG}{\mathcal{G}}

\newcommand{\cX}{\mathcal{X}}
\newcommand{\cY}{\mathcal{Y}}
\newcommand{\cZ}{\mathcal{Z}}

\theoremstyle{plain}
\newtheorem{theorem}{Theorem}

\theoremstyle{definition}
\newtheorem{definition}{Definition}

\newtheorem{protocol}{Protocol}
\newtheorem{task}{Task}
\newtheorem*{task*}{Task T}

\begin{document}

\title{Theories with no superluminal signaling have greater information-processing power
than theories with no superluminal causation}

\author{V. Vilasini}
\email{vilasini@inria.fr}
\affiliation{Université Grenoble Alpes, Inria, Grenoble 38000, France}
\affiliation{Institute for Theoretical Physics, ETH Zurich, 8093 Z\"{u}rich, Switzerland}
\author{Roger Colbeck}
\email{roger.colbeck@kcl.ac.uk}
\affiliation{Department of Mathematics, King's College London, Strand, WC2R 2LS, UK}
\affiliation{Department of Mathematics, University of York, Heslington, York YO10 5DD, UK}

\date{\today}

\begin{abstract}
A central goal in the foundations of physics is to understand the structure of physical theories, such as quantum theory, from physical principles. This is often explored by considering various information-theoretic principles. Here, we initiate a similar approach considering relativistic causality principles. No superluminal causation (NSC) and no superluminal signalling (NSS) are distinct relativistic principles, requiring, respectively, that causal influence/the ability of agents to signal are within the future lightcone. After formalizing their distinction, we investigate how well theories constrained by NSC and NSS perform in a task that involves generating non-classical correlations. We find a spacetime configuration in which this task cannot be achieved in any theory (classical, quantum, or post-quantum) satisfying NSC. However, we show that theories violating NSC but satisfying NSS can perfectly achieve the task. 
We give a protocol that would, in a world allowing superluminal causation, enable its operational certification without violating NSS, in general spacetimes. In the case of $(1+1)$D Minkowski spacetime, the task remains achievable in a configuration where measurement outcomes occur arbitrarily earlier in time than the settings, allowing a new form of certifiable retrocausality without violating NSS. We illustrate our results by linking two different types of non-classical post-quantum resources: PR-boxes and jamming. Our work offers insights into the role of different relativistic causality principles in fundamental physics and paves the way for characterising the information-theoretic structure of theories obeying such principles.
\end{abstract}

\maketitle

\ifarxiv\section{Introduction}\label{sec:intro}\else\noindent{\it Introduction.}|\fi Why does the physical theory describing the world give rise to certain phenomena but not others, and how would this change in another theory?  These are two key questions in the foundations of physics.  One way to explore these questions is to characterize classes of theories constrained by various principles. This has led to candidate reformulations of quantum theory based on information-theoretic principles (see e.g., \cite{Hardy2001,Chiribella2011,Masanes2011,Muller2021}). Here, instead, we focus on principles from relativistic causality.

One such principle is no superluminal causation (NSC), which limits how information can flow between different spacetime events in an underlying theory. The principle of no superluminal signalling (NSS) is a distinct yet related relativistic principle, constraining the ability of agents at different spacetime locations to communicate through interventions. These principles are distinct because causation can be present in an underlying theory without observable signalling between agents~\cite{Wood2015, Barrett2020A,VilasiniColbeckPRA}, thereby leaving open the possibility of violating NSC without violating NSS.

Although often conflated, a complete understanding of the quantum-relativity interface requires these principles to be formulated and studied independently. We do so within a recent general framework~\cite{VilasiniColbeckPRA,VilasiniColbeckPRL} developed by us and investigate the implications of NSC and NSS for information processing, considering the task of generating non-classical correlations\footnote{In this work by non-classical correlations we mean those that violate a Bell inequality.} (such correlations are a requirement for other tasks such as device-independent QKD~\cite{BHK,PABGMS} or randomness expansion~\cite{ColbeckThesis,CK2,PAMBMMOHLMM}). We examine the task in two spacetime configurations and show that in the first it is impossible in any \emph{classical} theory constrained by NSC while in the second, it is impossible in \emph{any} theory (classical or non-classical) constrained by NSC. This sets the stage for exploring whether these tasks may be possible in a theory that violates NSC but is constrained by the weaker principle of NSS. 

We emphasise that our motivation for considering violations of NSC is to understand why NSC should be used as a physical principle in addition to NSS. In particular, we are not suggesting that NSC is violated in nature and our results, regarding the implications of NSC violations, constitute good reasons why it should hold.

  

In the context of NSS in tripartite Bell scenarios, jamming non-local theories were proposed in~\cite{Grunhaus1996}. These theories allow a party (e.g., Bob) to influence the correlations between the measurement outcomes of space-like separated parties (e.g., Alice and Charlie) without affecting the individual outcomes detectably. It has been suggested~\cite{Grunhaus1996, Horodecki2019} that such theories do not allow superluminal signalling in Minkowski spacetime, but this was not proven within a formal theoretical framework, and the role of other relativistic causality principles, such as NSC, has remained little understood.

We revisit the aforementioned task in jamming theories, showing that here it can be achieved by violating NSC while satisfying NSS. We do so through a protocol that enables Alice and Charlie to generate maximally non-classical Popescu-Rohrlich (PR) box~\cite{Tsirelson,PR1994} correlations by sending classical information to Bob if Bob is able to jam their correlations. These results hence highlight an operational difference between NSC and NSS in a manner applicable to classical and non-classical theories. As a note of interest at the end of the paper we also discuss the speed of generating non-classical correlations via jamming in different scenarios, showing that in some cases this is possible even when measurement outcomes occur arbitrarily earlier than the corresponding measurement settings in some reference frame. This suggests a new form of retrocausality (backwards-in-time causal influences) that can be operationally certified in such a theory, using our protocol, in $(1+1)$D Minkowski spacetime.

Previous retrocausal models proposed for explaining Bell correlations (see e.g., \cite{Price_2008,Baumeler2018}) lack the same verifiability because non-classical causal models that only involve forward in time causation can also explain the same operational predictions (see e.g., \cite{Wiseman2015,Wood2015,Allen2017,Henson2014,Barrett2020A}). Therefore, our results can be seen as a stronger argument to reject theories that violate NSC even if NSS holds. 
In particular, such an argument rules out (post-quantum) theories with jamming correlations \cite{Grunhaus1996} or so-called relativistically causal correlations \cite{Horodecki2019}.

\ifarxiv\section{Jamming non-local correlations}\label{sec:jamming}\else\medskip

\noindent{\it Jamming non-local correlations.}|\fi
Consider a tripartite Bell scenario where Alice, Bob and Charlie measure some shared system $\Lambda$ using freely chosen settings $A$, $B$ and $C$ to obtain outcomes $X$, $Y$ and $Z$ respectively, in a spacetime configuration where their measurements are pairwise space-like separated. The standard tripartite non-signalling conditions on the correlations $P(XYZ|ABC)$ require that the outcomes of any set of parties are independent of the settings of the complementary set of parties, in particular $P(XZ|ABC)=P(XZ|AC)$ holds. 

If we are guaranteed that the joint future of the spacetime locations of $X$ and $Z$ is contained in the future of the spacetime location of $B$, then~\cite{Grunhaus1996} propose that $P(XZ|ABC)=P(XZ|AC)$ is no longer relevant for avoiding superluminal signalling since to access both $X$ and $Z$ one needs to be in the joint future of the spacetime locations of $X$ and $Z$, and hence in the future of the location of $B$. This leads to the concept of jamming non-local correlations, which allows Bob's setting, $B$, to be jointly but not individually correlated with Alice and Charlie's outcome, allowing Bob to \emph{jam} the correlations between Alice and Charlie depending on his setting. For the present paper, we consider $B$ jamming $X$ and $Z$, in which case we define correlations to be \emph{jamming} if they satisfy the conditions

\begin{subequations}
  \begin{align}
        P(XZ|B)&\neq P(XZ) \label{eq: jamming1}\\
        P(X|B)&=P(X)\label{eq: jamming2}\\
        P(Z|B)&=P(Z).\label{eq: jamming3}
\end{align}  
\end{subequations}

\ifarxiv\section{Causal structures and affects relations\label{sec:affs}}\else\noindent{\it Causal structures and their compatibility with spacetime.}|\fi
Following the framework of~\cite{VilasiniColbeckPRA}, we describe information-theoretic causal structures (henceforth simply referred to as causal structures) through an approach based on causal modelling and inference\footnote{Causal modelling and causal inference were originally developed in classical statistics \cite {Pearl2009} and recently generalised to quantum and other non-classical theories (e.g., \cite {Henson2014, Barrett2020A}).}. Here, \emph{causal structures} are modelled as directed graphs with directed edges $\longrsquigarrow$ denoting causation, and each node can either be an observed node (which corresponds to a classical variable such as a measurement setting or outcome) or an unobserved node (which can be a classical, quantum or a post-quantum system). The set of all observed nodes is denoted as $N_{\mathrm{obs}}$.

In a classical causal model over a causal structure $\mathcal{G}$, all nodes correspond to classical variables and the model is obtained by specifying, for each node $N$, a function $f_N$ that maps the values of the random variables that are parents of $N$ to the possible values of $N$ or, if $N$ is parentless in $\mathcal{G}$, a probability distribution $P(N)$. In quantum causal models, the deterministic functions are replaced with quantum channels, and in post-quantum theories, they are replaced with maps that describe the allowed transformations of that theory. We refer to these as the \emph{causal mechanisms} of the theory; they give meaning to the causal relation $\longrsquigarrow$ as a flow of information according to the rules of the theory.

We can associate a probability distribution, $P_{\mathcal{G}}(N_{\mathrm{obs}})$, with the observed nodes of $\mathcal{G}$, the set of allowed distributions depending on the underlying theory in general. The \emph{d-separation property} relates graph separations in $\mathcal{G}$ to corresponding conditional independences in $P_{\mathcal{G}}(N_{\mathrm{obs}})$ (see \ifarxiv Appendix~\ref{appendix: framework} \else the Supplemental Material \fi for more detail). All (possibly non-classical) causal models defined on directed acyclic graphs $\mathcal{G}$ are known to satisfy this property~\cite{Pearl2009, Tucci, Henson2014}, and some cyclic ones do too~\cite{Bongers2021, VilasiniColbeckPRA}. 
In a causal model, a variable $X$ is said to be freely chosen if it has no parents (i.e., no incoming arrows).

As well as accounting for correlations through $P_{\mathcal{G}}(N_{\mathrm{obs}})$, the framework also allows for interventions, which are needed to distinguish correlation and causation. An intervention takes a set $X$ of observed nodes in a causal structure $\mathcal{G}$ and associates them with an alternative causal structure $\mathcal{G}_{\mathrm{do}(X)}$, obtained from $\mathcal{G}$ by removing all incoming edges to $X$. This captures that the intervention on $X$ is freely chosen, i.e., independent of prior causes of $X$. For two disjoint subsets $X$ and $Y$ of observed nodes, we say that $X$ \emph{affects} $Y$ if there exists a value $x$ of $X$ such that
\begin{equation}
    P_{\mathcal{G}_{\mathrm{do}(X)}}(Y|X=x)\neq P_{\mathcal{G}}(Y).
\end{equation}
This encapsulates the idea that the probability of $Y$ can be changed by freely intervening on $X$. $X$ affects $Y$ captures that an agent can signal from $X$ to $Y$ using interventions on $X$, and this implies that $X$ is a cause of $Y$. However, the converse is not true, i.e., it is possible to have \emph{causation without signalling}. A common example is that of jamming, where~\eqref{eq: jamming2} and~\eqref{eq: jamming3} (together with the free choice of $B$, modelled as a parentless node) mean that $B$ cannot individually affect $X$ or $Z$, but there must be causation from $B$ to at least one of $X$ or $Z$, so that~\eqref{eq: jamming1} holds (see also Protocol \ref{protocol: jamming}).

An affects relation $X$ affects $Y$ as defined above captures that an agent with access to the variables $X$ can signal to an agent with access to variables $Y$ through interventions on $X$. More generally, 
we can consider signalling possibilities given that some interventions have been performed on another set $Z$ of variables. Signalling from $X$ to $Y$ given interventions on $Z$ is captured by a \emph{higher order affects relation}, $X$ affects $Y$ given do$(Z)$, which was introduced in \cite{VilasiniColbeckPRA} (see \ifarxiv Appendix~\ref{appendix: framework} \else the Supplemental Material \fi for the full definition). Higher order affects relations have a rich structure and are crucial for fully formalising the principle of no superluminal signalling (NSS) in a spacetime in these general scenarios \cite{VilasiniColbeckPRA, Grothus2024}. The formalisation of the NSC and NSS relativistic principles in our framework is presented in the next section.

\section{Embedding causal models in spacetime: NSC and NSS principles}

Causal models can be embedded in spacetime by assigning to each observed node $X$, a spacetime location, which leads to a corresponding spacetime random variable $\cX$. In our framework, spacetime is modelled as a partially ordered set $\mathcal{T}$ where $P\prec Q$, $P\succ Q$ and $P\not\prec \not\succ Q$ for $P,Q\in \mathcal{T}$ represent $P$ being in the future lightcone of, past lightcone of, and space-like separated to $Q$, respectively. An embedding $\mathcal{E}$ of a causal model with observed nodes $N_{\mathrm{obs}}$ (or a set $\mathscr{A}$ of affects relations over a set $N_{\mathrm{obs}}$ of variables) simply assigns a location in $\mathcal{T}$ to each random variable $X\in N_{\mathrm{obs}}$, thereby generating an ordered random variable (ORV) $\cX:=(X,O(X))$, where $O(X)\in \mathcal{T}$. ORVs are simply an abstraction of spacetime random variables, to general partially ordered sets. Moreover, the inclusive future of an ORV (i.e., points in the future lightcone of $\cX$ including the location of $\cX$)  is then given as

\begin{equation}
    \overline{\mathcal{F}}(\cX):=\{P\in \mathcal{T}|P\succeq O(X)\}.
\end{equation}

We can now formalise the two relativistic causality principles considered in this paper: no superluminal causation (NSC) and no superluminal signalling (NSS). More precisely, NSC (/NSS) imposes that when we embed an informational protocol (which can be equivalently described as a causal model) in spacetime by promoting the relevant variables of the protocols to spacetime variables, then there is no causal influence (/signalling) from one spacetime event to another event outside its future lightcone.

Within this general formalism, the NSC and NSS principles can be defined as follows.

\begin{definition}[NSC theories]
\label{definition: NSC}
Given a causal model over a causal structure $\mathcal{G}$ and an embedding $\mathcal{E}$ of the causal model in a spacetime $\mathcal{T}$, we say that the embedded model \emph{satisfies NSC} if whenever $X\longrsquigarrow Y$ ($X$ is a direct cause of $Y$) for any two nodes $X$ and $Y$ in $\mathcal{G}$, then $\cX\prec \cY$ in the spacetime embedding. An \emph{NSC theory} only allows for causal models of this type in any given spacetime embedding. A classical NSC theory is one that restricts all unobserved nodes of its causal models to be classical systems, otherwise it is called a non-classical NSC theory.
\end{definition}

The NSS principle is captured by the following compatibility condition between (higher-order) affects relations of the model and the spacetime embedding (originally proposed in \cite{VilasiniColbeckPRA}), noting that higher-order affects relations capture the general way of signalling in such scenarios.

\begin{definition}[Compatibility of a set of affects relations with an embedding in a partial order \cite{VilasiniColbeckPRA}]
\label{definition: compat}
Let $\mathcal{S}$ be a set of ORVs formed by embedding a set of RVs $S$ in a spacetime $\mathcal{T}$ with embedding $\mathscr{E}$. Then a set of affects relations $\mathscr{A}$ is said to be \emph{compatible} with the embedding $\mathscr{E}$ if whenever $X$ affects $Y$ given $\mathrm{do}(Z)$ belongs to $\mathscr{A}$, and is irreducible with respect to the affects relations in $\mathscr{A}$, then $\overline{\mathcal{F}}(\cY)\bigcap \overline{\mathcal{F}}(\cZ)\subseteq \overline{\mathcal{F}}(\cX) $ with respect to $\mathscr{E}$. 
\end{definition}

Since the affects relation $X$ affects $Y$ given $\mathrm{do}(Z)$ allows an agent with access to $X$ to signal to an agent with access to $Y$ and $Z$, the above compatibility condition captures the idea that in order to avoid superluminal signalling, we must require the affects relation to be embedded in spacetime such that $\overline{\mathcal{F}}(\cY)\bigcap \overline{\mathcal{F}}(\cZ)$ (which is where $Y$ and $Z$ can be jointly accessed) is fully contained in $\overline{\mathcal{F}}(\cX)$ (which is where $X$ can be accessed).

\begin{figure*}[ht!]
 \centering
\subfloat[\label{fig:JamPRsptime1}]{\includegraphics[scale=0.9]{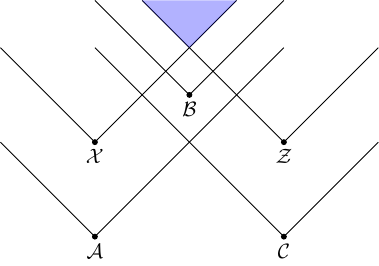}}\qquad\qquad \subfloat[\label{fig:JamPRsptime2}]{\includegraphics[scale=0.7]{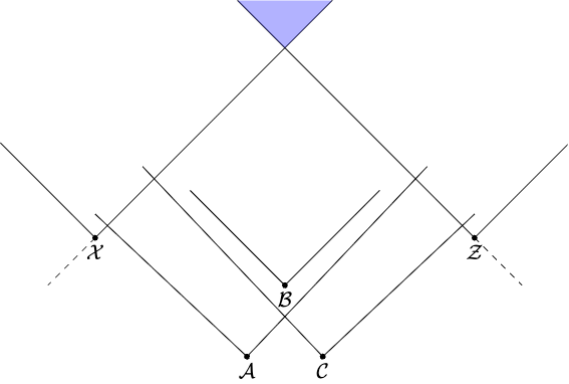}}
    \caption{Spacetime embeddings for the variables $A$, $C$, $X$ and $Z$ associated with (a) Task~\ref{task: task1} and (b) Task~\ref{task: task2}. 
    The definition of the tasks as well as the impossibility results of Theorem~\ref{theorem: impos} allow all possible spacetime embeddings of $B$. The strategy that achieves the task in jamming theories (Theorem~\ref{thm:JPR}) requires the additional variable $B$ to be embedded such that it is in the future of both $A$ and $C$, has the joint future of $X$ and $Z$ contained in its future, but does not have $X$ or $Z$ in its future (which holds as drawn in the figure).
}
    \label{fig: jammingPR}
\end{figure*}
\begin{definition}[NSS theories]
\label{definition: NSS}
    Given a causal model over a causal structure $\mathcal{G}$ and an embedding $\mathcal{E}$ of the causal model in a spacetime $\mathcal{T}$, we say that the embedded model \emph{satisfies NSS} if the set $\mathscr{A}$ of all affects relations of the causal model is compatible with the embedding $\mathcal{E}$ in the spacetime $\mathcal{T}$. An \emph{NSS theory} only allows for causal models of this type in any given spacetime embedding. A classical NSS theory is one that restricts all unobserved nodes of its causal models to be classical systems, otherwise it is called a non-classical NSS theory.
\end{definition}

Notice that since signalling between two variables, captured by an affects relation $X$ affects $Y$, implies that $X$ causes $Y$, it follows that NSC implies NSS in our formulation of these principles.

\section{Information processing in NSC and NSS theories}
\ifarxiv\subsection{Tasks that are impossible in theories with NSC}\else\smallskip \noindent{\it Tasks that are impossible in theories with NSC}|\fi 
We consider a primary task $T$ relating to the generation of non-classical correlations, and define two sub-tasks \ref{task: task1} and \ref{task: task2} by instantiating $T$ in two different classes of spacetime configurations. 

\begin{task*}
 The task involves two agents, Alice and Charlie. Each agent makes an input at some spacetime location, and receives a corresponding output variable at another spacetime location (which need not be in the future lightcone of the input variable). The input variables of Alice and Charlie are $A$ and $C$, while their output variables are $X$ and $Z$ respectively. The outcomes may be produced by interacting with a shared system $\Lambda$.  The task is for Alice and Charlie to use freely chosen inputs to generate non-classical correlations i.e., a distribution $P(XZ|AC)$ that cannot be expressed as
    \begin{equation}
\label{eq: classicalP}
    P(XZ|AC)=\sum_{V}P(V)P(X|A V)P(Z|C V)
\end{equation}
for any random variable $V$. Alice and Charlie may collaborate with a third agent Bob with whom they may share systems. 
\end{task*}
So far, there are no constraints on communication between Alice and Charlie and $T$ can be trivially achieved using classical communication. 

\begin{task}
\label{task: task1} 
Here, Alice and Charlie are required to achieve task $T$ given that variables $A$, $C$, $X$ and $Z$ are embedded in an acyclic spacetime $\mathcal{T}$ such that all four variables are spacelike separated except that $\cA\prec \cX$ and $\cC\prec \cZ$ (Figure~\ref{fig:JamPRsptime1}). 
\end{task}

\begin{task}
\label{task: task2} 
Here, Alice and Charlie are required to achieve task $T$ given that $A$, $C$, $X$ and $Z$ are embedded in an acyclic spacetime $\mathcal{T}$ such that $\overline{\mathcal{F}}(\cX)\cap \overline{\mathcal{F}}(\cZ)\subseteq \overline{\mathcal{F}}(\cA)\cap \overline{\mathcal{F}}(\cC)$ holds, 
but $\cA\not\prec \cX$ and $\cC\not\prec\cZ$ (Figure~\ref{fig:JamPRsptime2}). 
\end{task}

At this point the reader may be concerned about the physicality of the embedding of Figure~\ref{fig:JamPRsptime2} because, in general, if the input $A$ is a cause of the output $X$, we would usually expect $X$ to be in the future lightcone of $A$. However, in this work we are explicitly considering going beyond these usual notions by allowing superluminal causation (violating the NSC principle), provided it does not lead to superluminal signalling (the NSS principle still holds). The main goal of our work is to operationally distinguish between these distinct relativistic principles, hence it is necessary to consider theories violating NSC without violating NSS.

Then, the following theorem shows that there are limitations to achieving these tasks in theories restricted by the principle of no superluminal causation. We state informal versions of our theorems here, formal versions with proofs can be found in \ifarxiv Appendix~\ref{app:proof}\else the Supplemental Material\fi. 

\begin{restatable}{theorem}{Impossibility} (Informal)
\label{theorem: impos}
  Task~\ref{task: task1} is impossible to achieve in any classical theory without superluminal causation, while Task~\ref{task: task2} is impossible to achieve in any theory (classical or otherwise) without superluminal causation. This holds even when allowing for arbitrary collaboration strategies involving additional agents. 
\end{restatable}

The statement for Task~\ref{task: task1} is in line with what one would expect from Bell's theorem~\cite{Bell, Wood2015, Wiseman2015}. On the other hand, for Task~\ref{task: task2}, we show a stronger statement, that in a theory with NSC where $A$ and $C$ are freely chosen, it is impossible to generate any correlations between $\{A,C\}$ and $\{X,Z\}$ in the spacetime configuration of Task~\ref{task: task2} (Figure~\ref{fig:JamPRsptime2}) i.e., $P(XZ|AC)=P(XZ)$ always holds in such a theory. This implies the above theorem. 

\begin{figure}
    \centering
\includegraphics[scale=1.0]{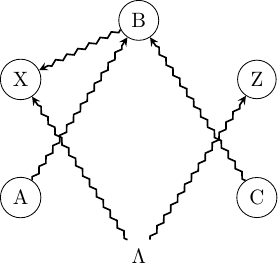}
    \caption{The causal structure corresponding to the causal model associated with Protocol~\ref{protocol: jamming}. The protocol combines the simplest form of jamming of the outputs $X$ and $Z$ by $B$ (whose causal model was given in \cite{VilasiniColbeckPRL}) and adds the dependence of $B$ on the inputs $A$ and $C$ which is given by the communication from Alice and Charlie to Bob in the protocol. The causal model is compatible with both the spacetime embeddings of Figure~\ref{fig: jammingPR}, i.e., satisfies the NSS principle relative to both these embeddings.  However, this manifestly violates the NSC principle since the causal arrow from $B$ to $X$ will flow outside the lightcone in those spacetime embeddings. 
    }
\label{fig: JamPRcausal}
\end{figure}

\ifarxiv\subsection{Achieving the tasks in theories with NSS}\else\smallskip \noindent{\it Achieving the tasks in theories with NSS}|\fi The following theorem shows that the above tasks which are impossible in a theory with NSC become possible in one with NSS. 
\begin{restatable}{theorem}{JammingPR}\label{thm:JPR} (Informal)
    Tasks~\ref{task: task1} and~\ref{task: task2} can both be achieved in a theory that realises jamming correlations, without violating NSS and using the assistance of an additional agent, Bob. Alice and Charlie only send classical communication to Bob, and exchange no communication otherwise. In particular, the maximally non-classical, PR box correlations can be generated in this manner.
\end{restatable}

We now describe a protocol that achieves Tasks~\ref{task: task1} and~\ref{task: task2}. The causal structure associated with this protocol is illustrated in Figure~\ref{fig: JamPRcausal}.

\begin{protocol}
\label{protocol: jamming}
Consider an additional agent Bob (associated with a variable $B$) who can jam the correlations between Alice and Charlie's output variables $X$ and $Z$, through his choice of $B$ (cf. Section~\ref{sec:jamming}). All variables are taken to be binary. Alice and Charlie communicate their classical inputs $A$ and $C$ to Bob, who sets $B=A.C$. Then through jamming, Bob can ensure that $B$ determines the parity of $X$ and $Z$ without being correlated individually with $X$ or $Z$, i.e., $B=X\oplus Z$. The result is $X\oplus Z=A.C$ which correspond to PR-box correlations $P(XZ|AC)$.
\end{protocol}

The jamming required in this protocol can be achieved through a classical common cause $\Lambda$ influencing $X$ and $Z$ as well as a superluminal causal influence from $B$ to $X$ or $Z$ \cite{VilasiniColbeck_Jamming}. The common cause $\Lambda$ must remain inaccessible to avoid the possibility of superluminal signalling using this influence. In \ifarxiv Appendix~\ref{app:proof}\else the Supplemental Material\fi, we show that there is a causal model for this protocol, defined on the causal structure of Figure~\ref{fig: JamPRcausal}, whose affects relations can be compatibly embedded in the spacetime configurations of Figures~\ref{fig:JamPRsptime1} and~\ref{fig:JamPRsptime2}. This proves that the protocol satisfies NSS. 

By modifying the associated causal model other non-classical correlations are also possible as explained in \ifarxiv Appendix~\ref{app:proof}\else the Supplemental Material\fi. Moreover, there is a natural connection between our protocol with jamming and non-local hidden variable theories, see \ifarxiv Section~\ref{app:NLHV} of the Methods \else the Supplemental Material \fi for details.

\ifarxiv\subsection{Certifying violation of NSC without violation of NSS}\else\smallskip \noindent{\it Certifying violation of NSC without violation of NSS}|\fi We have seen that strictly more can be achieved in theories with NSS than those with NSC. We now ask whether it would be possible to certify the presence of a superluminal causal influence in theories with NSS that achieve the tasks, as this would operationally establish that the violation of NSC is the relevant resource involved.

In Task~\ref{task: task1}, there is no theory-independent way to verify the superluminal causal influence because the task is possible without superluminal causal influence in quantum and post-quantum non-classical theories, i.e., achieving the tasks could be because of non-classicality rather than a violation of NSC.

By contrast, in Task~\ref{task: task2}, we can certify the existence of the superluminal causal influence in a theory-independent manner. In any theory without superluminal causation, we have shown that $P(XZ|AC)=P(XZ)$ holds in Task~\ref{task: task2}. By repeating a given setup and collecting statistics, if access to $A$ and $C$ enables $X$ and $Z$ to be jointly predicted better than without $A$ and $C$, then this certifies superluminal causal influence, irrespective of the underlying theory. Our setup with jamming described in Protocol~\ref{protocol: jamming} would enable such a certification since it allows the parity $X\oplus Z$ to be perfectly predicted given $A$ and $C$, even when the parity cannot be predicted better than a random guess when $A$ and $C$ are not given. 

In the next section we discuss an interesting consequence of our protocol in terms of how quickly two separated parties can generate non-local correlations between themselves.

\ifarxiv\subsection{Speed of creation of non-classical correlations}\else\smallskip\noindent{\it Speed of creation of non-classical correlations.}|\fi Consider again the task of generating non-classical correlations $P(XZ|AC)$ between Alice and Charlie. For simplicity, we discuss the case of $(1+1)$-dimensional Minkowski spacetime. Unless explicitly stated otherwise, the ideas here generalise to higher dimensions as well. Suppose that Alice and Charlie remain stationary at the spatial locations $x_A$ and $x_C>x_A$ respectively while carrying out their measurements. In a theory with NSC, any information travelling between the locations $x_A$ and $x_C$ that would correlate the variables $X$ and $Z$ produced there can at best travel at the speed of light $c$ (even if the communication of this information was mediated by other parties). Therefore the minimum time required to establish any non-classical correlations between Alice and Charlie if they do not have any pre-shared non-classical resources would be $t_{{\rm NSC}}=\frac{x_C-x_A}{c}$\ifarxiv{}\else{~}\fi\footnote{In the common rest frame of the parties, since we have taken
  them to be stationary at their respective locations.}.

On the other hand, consider the causal model of Protocol~\ref{protocol: jamming} embedded in spacetime as per Figure~\ref{fig:JamPRsptime1} where Alice's setting and outcome are associated with the same spatial location and similarly for Charlie, but the timing is such that the measurements are space-like separated. Our results imply that in a theory that allows jamming of space-like separated correlations, 
Alice and Charlie can establish non-classical correlations (such as the PR-box correlations) in half this time i.e., $t_{{\rm jam}}=\frac{x_C-x_A}{2c}$. 
They can do so with the help of a third party Bob situated half way between Alice and Charlie, who can receive $A$ and $C$ from them (which takes $t_{{\rm jam}}$ amount of time) and immediately jam their outputs to produce the required PR box correlations $X\oplus Z=A.C$. Thus non-classical correlations can be created twice as fast in such theories.

 If we allow the outputs $X$ and $Z$ to be produced at different spatial locations $x_X\leq x_A$ and $x_Z\geq x_C$, then in theories with NSC, more time would be required for information to travel from the input of one party to the output of the other and we would have $t_{{\rm NSC}}\geq \frac{x_C-x_A}{c}$. However, a surprising situation arises in jamming theories in $(1+1)$D Minkowski spacetime. It is possible to establish non-classical correlations even with a negative $t_{{\rm jam}}$ i.e., where the outputs are produced at a much earlier time than the inputs (in some frame of reference). Referring to Figure~\ref{fig:JamPRsptime2}, we can move the RVs $X$ and $Z$ down along a light-like surface (dashed lines) such that they end up arbitrarily far apart in space and time but their joint future (blue region) remains unaltered and continues to be contained in the future of the ORV $\cB$, enabling Bob to jam them. Thus we can always find a reference frame in which $t_{{\rm jam}}$ can be arbitrarily negative indicating retrocausality. We note that this form of retrocausality is not possible in Minkowski spacetimes with higher spatial dimensions \footnote{That this is not possible in the $(2+1)$D case follows from
  Lemma~C.5 of \cite {VilasiniColbeck_Jamming}. It follows from this lemma that
  if the spatial coordinates of $\protect \mathcal {A}$, $\protect \mathcal
  {X}$ and $\protect \mathcal {Z}$ all lie on a line, then in any frame where
  $\protect \mathcal {X}$ and $\protect \mathcal {Z}$ are simultaneous, their
  joint future (blue region) can only be contained in the future lightcone of
  $\protect \mathcal {A}$ if the time coordinate of $\protect \mathcal {A}$ is
  less than that of $\protect \mathcal {X}$ and $\protect \mathcal {Z}$ (and
  similarly for $\protect \mathcal {C}$). Thus, in contrast to the case of
  $(1+1)$ dimensions, $\protect \mathcal {X}$ and $\protect \mathcal {Z}$
  cannot be moved back in time without introducing superluminal signaling.}.

 Thus theories satisfying NSS (such as jamming theories) can in general create non-classical correlations in a shorter time than in a theory constrained by the stronger principle of NSC. 
 While this statement is generally true, the effect can take a more drastic retrocausal form in $(1+1)$D Minkowski spacetime.

 \ifarxiv\section{Discussion}\else\smallskip\noindent{\it Conclusions and outlook.}|\fi We have studied the relationship between the (sometimes conflated) relativistic principles of no superluminal causation and no superluminal signalling. That these are distinct has been noted before in the context of Bell’s theorem which implies a gap between classical theories constrained by NSC vs.\ NSS. Here, we have discussed an information theoretic task that would, in principle, allow the two principles to be operationally distinguished regardless of whether the underlying theory is classical or non-classical.

Although no experimental evidence currently suggests that either NSC or NSS is violated by nature, understanding how they constrain the landscape of possible physical theories is valuable for the foundations of physics. Recent work~\cite{VilasiniColbeck_Jamming} demonstrates that jamming can lead to a violation of NSS in certain scenarios, contrary to previous claims~\cite{Grunhaus1996, Horodecki2019}, but even in scenarios where jamming obeys NSS, it can produce counterintuitive effects due to the certifiable violation of NSC. Since NSC is a fundamental relativistic principle, its certifiable violation in jamming theories, as shown by our protocol, supports arguments for the non-physicality of such theories~\cite{VilasiniColbeck_Jamming, Weilenmann2023}.

Our work provides a connection between jamming non-local correlations and other post-quantum correlations, such as PR boxes, while demonstrating that jamming theories in $(1+1)$D Minkowski spacetime permit a form of retrocausality that could in principle be operationally verified without violating NSS. The present work goes beyond our previous work~\cite{VilasiniColbeckPRL}, which demonstrated causal loops without superluminal signalling in $(1+1)$D, by establishing an operational gap between NSC and NSS even in theories having an acyclic causal structure and hence a well-defined ordering of agents' operations, while linking this to a novel form of certifiable retrocausality that can exist without invoking causal loops. This has wider implications in both physics and philosophy, since retrocausality has been proposed as a mechanism for Bell inequality violations (see, e.g.,~\cite{Standford_retrocausality,Wharton2020, Wharton2018, Price_2008} for a review) and our findings can help understanding information-processing gaps between theories with definite vs.\ indefinite causal order~\cite{Hardy2005, Oreshkov2012, Chiribella2013} from the perspective of relativistic principles.

Furthermore, one can consider the consequences for theories with NSS arising from the possibility of achieving Task~\ref{task: task2}. If achieving Task~\ref{task: task2} is deemed unphysical, then theories allowing superluminal causation while respecting NSS must eliminate protocols like ours using additional physical principles. Non-local hidden variable theories, which posit extra variables to explain quantum correlations, are one example (see \ifarxiv Appendix~\ref{app:NLHV} \else the Supplemental Material \fi for further details on how they relate to our results). On the other hand, if Task~\ref{task: task2} is considered physical, the violation of relativistic principles such as NSC and no retrocausality could become certifiable through our protocol even in theories with NSS.

Future research directions include exploring the information processing capabilities and limitations of theories with these principles and related principles such as no causal loops (e.g., as formalized in~\cite{VilasiniColbeckPRA, VilasiniColbeckPRL}) more widely. For instance, the achievability of Task~T is a prerequisite for applications in device-independent cryptography~\cite{BHK,PABGMS,ColbeckThesis,CK2,PAMBMMOHLMM}, and the ability to generate PR box correlations has further implications for communication complexity~\cite{vD}. A wider comparison of these principles and their significance for information processing would further enhance our understanding of the foundations of physics and inform the discussion of how new physical theories might be formulated.

\ifarxiv{\acknowledgements}\else\smallskip
\noindent{\it Acknowledgements.}|\fi Part of this research was carried out when VV was a PhD candidate supported by a Scholarship from the Department of Mathematics at the University of York. VV also acknowledges support from an ETH Postdoctoral Fellowship, the ETH Zurich Quantum Center, the Swiss National Science Foundation via project No.\ 200021\_188541 and the QuantERA programme via project No.\ 20QT21\_187724 and the PEPR integrated project EPiQ ANR-22-PETQ-0007 as part of Plan France 2030.


\appendix

\section{Further details of the causal modelling framework}
\label{appendix: framework}

In the main text, we have provided a rather non-technical overview of the causality framework of \cite{VilasiniColbeckPRA, VilasiniColbeckPRL} which yields a formalism for backing our main results about information processing and relativistic causality principles. Here, we present further technical details of this framework which are relevant for the arguments of this paper. A first important concept is that of \emph{d-separation}, which is defined below.

\begin{definition}[Blocked paths]
Let $\mathcal{G}$ be a directed graph in which $X$ and $Y\neq X$ are nodes and let $Z$ be a set of nodes not containing $X$ or $Y$.  A path from $X$ to $Y$ (not necessarily directed) is said to be \emph{blocked} by $Z$ if it contains either $A\longrsquigarrow W\longrsquigarrow B$
with $W\in Z$, $A\longlsquigarrow W\longrsquigarrow B$ (a \emph{fork})
with $W\in Z$ or $A\longrsquigarrow W\longlsquigarrow B$ (a \emph{collider}) such that neither $W$ nor any descendant of $W$ belongs to $Z$, where $A$ and $B$ are arbitrary nodes in the said path between $X$ and $Y$.
\end{definition}

\begin{definition}[d-separation]
Let $\mathcal{G}$ be a directed graph in which $X$, $Y$ and $Z$ are disjoint
sets of nodes.  $X$ and $Y$ are \emph{d-separated} by $Z$ in
$\mathcal{G}$, denoted as $(X\perp^d Y|Z)_{\cG}$ 
if every path from a variable in $X$ to a variable in $Y$ is blocked by $Z$, otherwise, $X$ is said to be \emph{d-connected} with $Y$ given $Z$.
\end{definition}

D-separation is itself purely a property of the graph. What we call the d-separation property, then links d-separation in the causal structure $\mathcal{G}$ to conditional independences over the set $N_{\mathrm{obs}}$ of observed nodes of $\mathcal{G}$.

\begin{definition}[d-separation property]
\label{def: d-sep-prop}
Let $\mathcal{G}$ be a causal structure associated with the set $N_{\mathrm{obs}}$ of observed nodes, a distribution $P_{\mathcal{G}}(N_{\mathrm{obs}})$ is said to satisfy the \emph{d-separation property} with respect to $\mathcal{G}$ if and only if the following holds, where $X$, $Y$ and $Z$ are three disjoint subsets of $N_{\mathrm{obs}}$ with $X$ and $Y$ being non-empty.
\begin{align}
    \begin{split}
        (X&\perp^d Y|Z)_{\mathcal{G}}\\ &\Downarrow\\
        P_{\mathcal{G}}(XY|Z)&=P_{\mathcal{G}}(X|Z)P_{\mathcal{G}}(Y|Z).
    \end{split}
\end{align}
    
\end{definition}

Then a \emph{causal model} is specified by a directed graph $\mathcal{G}$ associated with a set of $N_{\mathrm{obs}}$ of observed nodes, together with a distribution observed $P_{\mathcal{G}}(N_{\mathrm{obs}})$ that satisfies the d-separation property with respect to $\mathcal{G}$. This provides a minimal and general definition of a causal model.

In the main text, we described how a classical, quantum or post-quantum causal model over a causal structure $\mathcal{G}$ can be obtained by specifying the causal mechanisms (the informational channels of the theory that we associate with the graph $\mathcal{G}$). The causal mechanisms allow us to compute the observed distribution $P_{\mathcal{G}}(N_{\mathrm{obs}})$. For instance, in quantum theory, the causal mechanisms correspond to quantum channels, and the observed distribution can be computed using the Born rule. It has been shown that for classical and non-classical causal models defined on directed acyclic graphs, the d-separation property holds for the resulting observed distribution~\cite{Henson2014, Barrett2020A}. More generally, our framework of~\cite{VilasiniColbeckPRA, VilasiniColbeckPRL} does not make assumptions about the causal mechanisms, but rather only assumes the d-separation property at the level of the observed distribution. This property can also be satisfied in several classical and non-classical models defined on cyclic graphs, making the framework more generally applicable to cyclic, non-classical causal structures. However for the present paper, only acyclic graphs will be relevant. 

In Section~\ref{sec:affs} we have reviewed the concept of \emph{affects relations} which capture the ability of agents to signal between sets of observed nodes through active interventions. Our formalism also introduces the concept of \emph{higher-order affects relations} to model more general interventions which captures the ability of agents to signal from $X$ to $Y$ when given information about interventions performed on a third set $Z$ of observed nodes. These are defined as follows.

\begin{definition}[Higher-order affects relation]
\label{definition:HOaffects}
Consider a causal model associated with a causal graph $\cG$ over a set $N_{\mathrm{obs}}$ of observed nodes and an observed distribution $P_{\cG}$. Let $X$ , $Y$, and $Z$ be three pairwise disjoint subsets of $N_{\mathrm{obs}}$, with $X$ and $Y$ non-empty. We say that $X$ affects $Y$ given do$(Z)$ if there exists values $x$ of $X$, $z$ of $Z$ such that
\begin{equation}
\label{eq: HOaffects}
   P_{\cG_{\mathrm{do}(X Z)}}(Y|X=x,Z=z)\neq P_{\cG_{\mathrm{do}(Z)}}(Y|Z=z),
\end{equation}
which we denote in short as $ P_{\cG_{\mathrm{do}(X Z)}}(Y|X,Z)\neq P_{\cG_{\mathrm{do}(Z)}}(Y|Z)$. When $Z= \emptyset$, we refer to this as a \emph{zeroth-order affects relation}.
\end{definition}

In general two affects relations $X_1$ affects $Y$ and $\{X_1,X_2\}$ affects $Y$ can carry the same information, this happens for instance when $X_2$ is irrelevant to the scenario (e.g., it may have no causal connections with $X_1$ or $Y$). In such a case, the affects relation emanating from the smaller set is the relevant one, both for causal inference (as it tells us that $X_1$ is a cause of $Y$) and for relativistic causality principles (as requiring $Y$ to be embedding in the future lightcone of $X_1$ and $X_2$ in light of the second affects relation may not be necessary for avoiding superluminal signalling, if $X_2$ does not add any information to the signal).  
Then the following definition captures whether an affects is reducible, which is important for ensuring that our compatibility condition is both a necessary and a sufficient condition for having no superluminal signalling when a causal model is embedded in spacetime.

\begin{definition}[Reducible and irreducible affects relations]
\label{definition: ReduceAffects}
If for all proper subsets $s_X$ of $X$, $s_X$ affects $Y$ given $\mathrm{do}(Z \tilde{s}_X)$, the affects relation $X$ affects $Y$ given $\mathrm{do}(Z)$ is said to be \emph{irreducible}. Otherwise, it is said to be \emph{reducible}.
\end{definition}

It is shown in \cite{VilasiniColbeckPRA} that for every reducible higher-order affects relation $X$ affects $Y$ given $\mathrm{do}(Z)$, there exists a proper subset $\tilde{s}_X$ of $X$ such that $\tilde{s}_X$ affects $Y$ given $\mathrm{do}(Z)$ i.e., $X$ affects $Y$ given $\mathrm{do}(Z)$ can be reduced to the affects relation $\tilde{s}_X$ affects $Y$ given $\mathrm{do}(Z)$.

\section{Proofs of all results}\label{app:proof}

We reproduce the informal statements of the theorems from the main text, before giving their formal versions and proofs.  Both results will refer to the following two spacetime embeddings of the variables $A$, $C$, $X$ and $Z$ modelling Alice and Charlie's inputs and outputs.

\begin{definition}[Embedding 1]
\label{def: emb1}
    $\cA\nprec \nsucc \cC$, $\cA\nprec \nsucc \cZ$, $\cX\nprec \nsucc \cC$, $\cX\nprec \nsucc \cZ$ (i.e., these pairs are space-like separated) and $\cA\prec \cX$, $\cC \prec \cZ$  (e.g., Fig.~1(a)).
\end{definition}

\begin{definition}[Embedding 2]
\label{def: emb2}
    $\cA\not\prec\not\succ \cX$, $\cC\not\prec\not\succ\cZ$, $\cA\not\prec\not\succ\cZ$ and $\cC\not\prec\not\succ\cX$  and $\overline{\mathcal{F}}(\cX)\cap \overline{\mathcal{F}}(\cZ)\subseteq \overline{\mathcal{F}}(\cA)\cap \overline{\mathcal{F}}(\cC)$ (e.g., Fig.~1(b)).
\end{definition}

Further we recall the following definition of non-classical correlations.
\begin{definition}[Non-classical correlations]
We say that $P(XZ|AC)$ is non-classical if it cannot be written as follows for any set of variables $V$,
\begin{equation}
\label{eq: classical}
    P(XZ|AC)=\sum_V P(X|AV)P(Z|CV)P(V).
\end{equation}    
\end{definition}

\subsection{Proof of Theorem 1}

\setcounter{theorem}{0}
\begin{restatable}{theorem}{Impossibility} (Informal)
\label{theorem: impos}
  Task 1 is impossible to achieve in any classical theory without superluminal causation, while  Task 2 is impossible to achieve in any theory (classical or otherwise) without superluminal causation. This holds even when allowing for arbitrary collaboration strategies involving additional agents. 
\end{restatable}

\setcounter{theorem}{0}
\begin{theorem}(Formal)
\label{theorem: impos_formal}
Let $A$, $C$, $X$ and $Z$ be random variables and $\Lambda$ be a classical or non-classical system. Further, let $S$ be an arbitrary set of random variables and systems. In any causal model over these variables/systems in which $A$ and $C$ are parentless, it is impossible to generate non-classical correlations $P(XZ|AC)$ in the following two cases:
\begin{enumerate}
    \item The causal model is classical and satisfies NSC with Embedding 1 (Definition~\ref{def: emb1}).
    \item The causal model (which may be classical or non-classical) satisfies NSC with Embedding 2 (Definition~\ref{def: emb2}).
\end{enumerate}

\end{theorem}
The formal version of this theorem is a statement about the observed random variables $A$, $C$, $X$ and $Z$ which are part of a causal model (potentially involving other variables and systems), and the embedding of this causal model in spacetime. To connect to the informal version, take $A$ and $C$ to play the role of Alice's and Charlie's settings and $X$ and $Z$ to be their outcomes. The conditions on the causal model and embedding capture the conditions of the tasks. For instance, the free choice of $A$ and $C$ is captured by taking them to be parentless in the causal model. 

\begin{proof}

    We split the proof into two parts, covering the statement for Embeddings~1 and~2 separately. We discuss the case without $S$, before extending the argument to cover such additional systems.\medskip

{\bf Impossibility for Embedding~1:} The distribution $P(XZ|AC)$ is the one relevant for the task. Since we are working in a classical theory, all the systems can be taken to be classical random variables. We now show that NSC in the spacetime embedding of Definition~\ref{def: emb1} implies that there exists a variable $V$ such that the correlations $P(XZ|AC)$ factorise as in Equation~\eqref{eq: classical}, i.e., that non-classical correlations cannot be generated.


For any random variable $\Lambda$, we can write $P(XZ|AC)=\sum_{\Lambda}P(XZ\Lambda|AC)$ and express this as follows using the rules of conditional probability.
\begin{equation}
\label{eq: proof_lemma1}
    P(XZ|AC)=\sum_{\Lambda}P(X|\Lambda Z AC)P(Z|\Lambda AC)P(\Lambda|AC).
\end{equation}


We now apply NSC and the fact that $A$ and $C$ are parentless to identify d-separations in any causal model embedded as per Definition~\ref{def: emb1}. This will yield corresponding conditional independences in the probability distribution, by the d-separation property (Definition~\ref{def: d-sep-prop}) and allow us to simplify Equation~\eqref{eq: proof_lemma1} to the required classical form.

\begin{enumerate}
\item Since $\cX$ and $\cZ$ are spacelike separated, NSC means there are no directed paths between $X$ and $Z$ in the causal model. First suppose $X$ and $Z$ have no common cause. This means paths between $X$ and $Z$ cannot contain a fork, and hence any paths between them must have a collider. Since $A$ and $C$ are parentless, the variable at any such collider cannot be $A$ or $C$ or have $A$ or $C$ as a descendant. Thus, $(X\perp^d Z|AC)$, and hence $P(XZ|AC)=P(X|AC)P(Z|AC)$. By a similar argument $(X\perp^d C|A)$, and symmetrically $(Z\perp^d A|C)$, so that $P(XZ|AC)=P(X|A)P(Z|C)$, which is a special case of~\eqref{eq: classical} with $V$ trivial.

Hence, we can assume $X$ and $Z$ have a common cause, $\Lambda$. NSC implies that $\Lambda$ must be embedded in the joint past of $\cX$ and $\cZ$ and that there cannot be any causal influences from $X$ or $Z$ to $\Lambda$. 
    This, together with $A$ and $C$ being parentless implies $(X\perp^d Z|\Lambda AC)$, i.e., $P(X|\Lambda Z AC)=P(X|\Lambda AC)$ by Definition~\ref{def: d-sep-prop}.
    \item As per above, let $\Lambda$ represent a variable embedded in the past light cone of $\cX$ and $\cZ$, without loss of generality. Since $A$ and $C$ are parentless, they could only be d-connected with $\Lambda$ if they had a directed path to $\Lambda$. Using $\cA \nprec \nsucc \cZ$ and $\cC \nprec \nsucc \cX$ together with NSC rules out such directed paths. Hence, $(\Lambda \perp^d AC)$, which implies the conditional independence $P(\Lambda|AC)=P(\Lambda)$. Note that if we were to instead consider a variable $\Lambda$ that is embedded in the future light cone of $\cA$ or $\cC$, then this $\Lambda$ could not be a common cause of $\cX$ and $\cZ$ due to NSC and cannot correlate them. Without loss of generality, such variables could be absorbed into $A$ or $C$ itself. 
    \item Since $\cX$ and $\cC$ are spacelike separated, NSC means there are no directed paths between $X$ and $C$ in the causal model. Since $C$ has no incoming arrows, all paths between $X$ and $C$ must contain a collider with central variable in the future of $C$. It follows that $(X\perp^d C|A \Lambda)$, and consequently $P(X|\Lambda AC)=P(X|\Lambda A)$. Symmetrically, we also have $(Z\perp^d A|C \Lambda)$ and hence $P(Z|\Lambda AC)=P(Z|\Lambda C)$.
\end{enumerate}

Plugging all of these conditional independences in Equation~\eqref{eq: proof_lemma1}, we obtain 
\begin{equation}
    P(XZ|AC)=\sum_{\Lambda}P(X|\Lambda A)P(Z|\Lambda C)P(\Lambda),
\end{equation}
which is precisely the condition for the correlation $P(XZ|AC)$ being classical (Eq.~\eqref{eq: classical}, with $\Lambda$ taking the role of $V$).

To extend the argument to the case of arbitrary additional systems $S$, consider the systems $S$ being described by some causal model satisfying the d-separation property. In all the above arguments, the d-separation criteria used in the proof follow from NSC and free choice of $A$ and $C$, and hence hold even when considering causal influences mediated by additional variables $S$. This proves the statement for Task~1.

\medskip

{\bf Impossibility for Embedding~2:} Since $\cX$, $\cZ$, $\cA$ and $\cC$ are all spacelike separated, NSC means that there can be no directed paths between any of the 4 variables in the causal model.  Furthermore, since $A$ and $C$ are parentless, any path between $AC$ and $XZ$ must contain a collider whose central variable is in the future of at least one of $A$ or $C$. It follows that $(XZ\perp^d AC)$, which implies $P(XZ|AC)=P(XZ)$, and hence that no non-classical correlations are possible (any distribution $P(XZ)$ can be written in the form $\sum_VP(V)P(X|V)P(Z|V)$).
\end{proof}

\subsection{Proof of Theorem 2}
\setcounter{theorem}{1}
\begin{restatable}{theorem}{JammingPR}\label{thm:JPR} (Informal)
    Tasks 1 and 2 can both be achieved in a classical theory that realises jamming correlations, without violating NSS and using the assistance of an additional agent, Bob. Alice and Charlie only send classical communication to Bob, and exchange no communication otherwise. In particular, the maximally non-classical, PR box correlations can be generated in this manner.
\end{restatable}

To show this we use the causal structure of Figure~\ref{fig: JamPRcausal}, in which $\Lambda$ is unobserved and all other nodes are observed, and all the nodes correspond to binary classical variables. In the causal model, we take $\Lambda$ to be uniformly distributed, $Z=\Lambda$, $B=A.C$ and $X=\Lambda\oplus B$. This is the jamming causal model presented in~\cite{VilasiniColbeckPRL} (which is defined on $X$, $Z$, $B$ and $\Lambda$) with the addition of the parentless nodes $A$ and $C$, which causally influence $B$. This leads to the PR box correlations $X\oplus Z=A.C$, as in Protocol~\ref{protocol: jamming}. Moreover, this model involves jamming correlations between $B$, $X$ and $Z$ since the conditions in Eqs.~\eqref{eq: jamming1}--~\eqref{eq: jamming3} are satisfied: $B=X\oplus Z$ ensures the first condition and $\Lambda$ being uniform ensures the last two.

[Note that a wider range of correlations can be generated by modifying the causal model to include a ``noise'' variable. For instance, defining $\Lambda=(E,F)$ where $E$ is uniformly distributed, we can have $X=E$, $B=A.C$ and $Z=E\oplus F\oplus B$ and, by varying the distribution over $F$, we can generate other (possibly quantum) correlations as well. The given model is asymmetric as there is only an influence of $B$ on $Z$ and not on $X$, but it can be symmetrised, for instance by including influences from $B$ to $X$ and $Z$ and considering the following functional dependences where we include an additional symmetrising variable in $\Lambda$: $\Lambda:=(E,F,G)$ with $E$ and $G$ uniform, $B=A.C$, $X=E\oplus(G\oplus 1).(F\oplus B)$, $Z=E\oplus G.(F\oplus B)$.]

\setcounter{theorem}{1}
\begin{theorem}(Formal)
  Consider the causal structure of Figure~\ref{fig: JamPRcausal} and the causal model described above. For each of Embeddings~1 and~2 (Definitions~\ref{def: emb1} and \ref{def: emb2}), there exist locations for the variables in the model compatible with the embedding, together with a location for the variable $B$ such that the embedded model satisfies NSS.
  \end{theorem}


Note that the causal model described captures what is needed for Protocol~\ref{protocol: jamming}.
  
To relate the given causal model to the protocol, $B$ plays the role of the jamming variable under the control of the agent Bob. Taking $B=A.C$ is possible because $A$ and $C$ are communicated to Bob. $B$ affects $\{X,Z\}$ (together with the absence of affects relations from $B$ to $X$ or $Z$) capture the ability of Bob to jam Alice and Charlie's correlations. 

\begin{proof}
In the given causal model, $A$ affects $B$. As $A$ is parentless, this affects relation is equivalent to $P(B|A)\neq P(B)$ which holds because $B=A.C$ and $C$ is not fixed to $0$. By the same argument, $C$ affects $B$. For NSS we hence require $\cB\succ \cA$ and $\cB\succ \cC$, which are equivalent to $\overline{\mathcal{F}}(\cB)\subseteq\overline{\mathcal{F}}(\cA)$ and $\overline{\mathcal{F}}(\cB)\subseteq\overline{\mathcal{F}}(\cC)$. We hence embed $B$ such that $\cB\succ\cA$ and $\cB\succ\cC$ (see Fig.~\ref{fig: jammingPR}), so these affects relations are compatible. 

Moreover, we also have $B$ affects $\{X,Z\}$ (since $P_{\mathcal{G}}(XZ)$ is uniform but $P_{\mathcal{G}_{\mathrm{do}(B)}}(XZ|B)$ is deterministic due to $B=X\oplus Z$). We hence need to embed $B$ such that $\overline{\mathcal{F}}(\cX)\cap \overline{\mathcal{F}}(\cZ)\subseteq\overline{\mathcal{F}}(\cB)$. Similarly we also have $\{A,C\}$ affects $\{X,Z\}$ so require $\overline{\mathcal{F}}(\cX)\cap \overline{\mathcal{F}}(\cZ)\subseteq\overline{\mathcal{F}}(\cA)\cap \overline{\mathcal{F}}(\cC)$. Both these affects relations are compatible with an embedding satisfying the conditions of Embeddings~1 and~2 with the addition of $\cB$ (an example is given in Fig.~\ref{fig: jammingPR}). 

Further, $B$ does not affect $X$ (since $P_{\mathcal{G}}(X)$ and $P_{\mathcal{G}_{\mathrm{do}(B)}}(X|B)$ are both uniform because $\Lambda$ is uniform) and similarly, $B$ does not affect $Z$. No further compatibility conditions are imposed by the affects relations of this causal model (for instance, the higher-order affects relation $A$ affects $B$ given do$(X)$ also holds but the compatibility condition for this relation, namely $\overline{\mathcal{F}}(\cB)\cap \overline{\mathcal{F}}(\cX)\subseteq\overline{\mathcal{F}}(\cA)$ is already implied by the compatibility condition for $A$ affects $B$, which is $\overline{\mathcal{F}}(\cB)\subseteq\overline{\mathcal{F}}(\cA)$), and hence the result is established.
\end{proof}
Note that the possible locations of $B$ in the embeddings depends on the spatial dimension, as a result of the requirement that $\overline{\mathcal{F}}(\cX)\cap \overline{\mathcal{F}}(\cZ)\subseteq\overline{\mathcal{F}}(\cB)$. In $(1+1)$-D, in a frame where $X$ and $Z$ are simultaneous, $B$ can be embedded later in time than $X$ and $Z$, while in higher dimensions, $B$ must be earlier in time than $X$ and $Z$ or simultaneous with them (see Lemma C.5 of~\cite{VilasiniColbeck_Jamming}).

\section{Relation to non-local hidden variable models}\label{app:NLHV}

In Theorem~2 we have shown that Tasks~1 and~2 can be achieved in jamming theories that violate NSC but not NSS. The protocol achieving the tasks highlights a link between the resources of jamming and the PR box, both of which have been proposed in different classes of post-quantum theories. Here, we discuss how the two tasks can also be achieved in other NSS theories without jamming. For example, we can slightly modify the causal model of our jamming protocol, by removing Bob's node $B$ in its causal structure (Fig.~\ref{fig: JamPRcausal}) and having $A$ and $C$ directly influence $X$. Then instead of the model $B=A.C$ and $X=\Lambda\oplus B$, we can directly set $X=\Lambda\oplus A.C$ to achieve the same correlations $P(XZ|AC)$ (in this case, PR box correlations). 

In the spacetime configurations for Tasks~1 and~2, the influence from $\cC$ to $\cX$ will be superluminal. As before, this model remains compatible with the spacetime i.e., does not lead to superluminal signalling in both the spacetime configurations, as long as $\Lambda$ remains unobserved.  In Task~1, this would correspond to an explanation of non-classical correlations in a Bell experiment in terms of a non-local hidden variable theory~\cite{Wood2015}. Moreover, this model can also be adapted analogously to the jamming case mentioned above, to generate other non-classical correlations, by adding a noise variable to $\Lambda$.

In the context of our discussion on the speed of creation of non-classical boxes, non-local hidden variable theories in principle allow for instantaneous creation of non-classical correlations,  
since the causal influence from $\cC$ to $\cX$ can be instantaneous. Furthermore, analogous to the case of jamming, the above model can also achieve arbitrarily negative times for creating non-classical correlations in the spacetime configuration of Task~\ref{task: task2} (Fig.~\ref{fig:JamPRsptime2}).

\end{document}